\documentclass[]{spie}  

\usepackage{amsmath,amsfonts,amssymb}
\usepackage{graphicx}
\usepackage[colorlinks=true, allcolors=blue]{hyperref}

\title{Principal component analysis of the Chandra ACIS gain}
\author[a]{Hans Moritz G\"unther}
\author[b]{\'Akos Bogd\'an}
\author[b]{Nick Durham }
\affil[a]{MIT Kavli Institute for Astrophysics and Space Research, Massachusetts Institute of Technology, Cambridge, MA 02139, USA}
\affil[b]{Smithsonian Astrophysical Observatory, Cambridge, MA 02138, USA}

\authorinfo{Send correspondence to H.M.G. (E-mail: hgunther@mit.edu)}

\begin{document}
\maketitle

\begin{abstract}
Up to 2020, the Chandra ACIS gain has been calibrated using the External Calibration Source (ECS). The ECS consists of an $^{\rm{55}}$Fe radioactive source and is placed in the ACIS housing such that all chips are fully illuminated. Since the radioactive source decays over time with a half-life of 2.7 years, count rates are becoming too low for gain calibration. Instead, astrophysical calibration sources will be needed, which do not fill and illuminate the entire field of view. Here, we determine the dominant spatial components of the gain maps through principal component analysis (PCA). We find that, given the noise levels observed today, all ACIS gain maps can be sufficiently described by just a few (often only one) spatial components. We conclude that illuminating a small area is sufficient for gain calibration. We apply this to observations of the astrophysical source Cassiopeia A. The resulting calibration is found to be accurate to 0.6\% in at least 68\% of the chip area, following the same definition for the calibration accuracy that has been used since launch.
\end{abstract}

\keywords{Chandra, calibration, gain, principal component analysis, ACIS, CCD}


\section{INTRODUCTION}
\label{sec:intro}
X-ray astronomy has been revolutionized by the advent of CCD detectors. With CCDs, we can determine the energy of every single incoming photon and thus extract an X-ray spectrum at any position of the image. In this way, spectra of many point sources, such as a stars in a young stellar cluster, or of different positions in an extended source, such as a supernova remnant, can be obtained in a single observation. However, CCDs do not directly output the photon energy. When an X-ray photon hits the CCD, it generates an electron cloud in the detector material. The number of electrons in that cloud is related to the photon energy. However, electrons can be lost if either the cloud spreads over too many pixels or if they are trapped on defects, causing a charge-transfer inefficiency (CTI)\cite{2002NIMPA.486..751T,GrantCTI}. After the electrons are read out, the signal needs to be amplified and processed electronically, until a PHA (Pulse Height Amplitude) value for a specific event is recorded. This PHA value is roughly linearly related to the photon energy, but the exact relation needs to be calibrated empirically using a source of photons of known energy. The gain changes over time, as the CCD ages, the particle contamination evolves over the solar cycles and electronic drift in the component occurs. Thus, this calibration needs to be repeated regularly; for Chandra every three to six months.

For the ACIS detector on the Chandra X-ray Observatory, this calibration has long relied on the so called ``external calibration source'' (ECS). The ECS consists of a $^{55}$Fe source and aluminum and titanium targets, thus emitting mostly in strong Al K$\alpha$, and Ti and Mn K$\alpha$ and K$\beta$ lines. The source is mounted such that it illuminates all ACIS chips uniformly while they are stowed during radiation belt passage.

Relying upon the ECS, the gain maps for each chip are currently constructed by fitting a number of regions independently. The region sizes (256 regions of $32 \times 128$ pixels each) were chosen to balance spatial resolution with ECS count statistics. One can see that the maps show consistent large scale structure (Figure~\ref{fig:explvar}). As the calibration source ages and count rates decline, fitting 256 regions independently is not possible any longer. Here, we present a scheme that makes use of the spatial structure in the gain maps.
The basic idea is to use PCA (Principal Component Analysis). Each observed gain map can be thought of as a vector with 256 features. PCA will find new bases vectors in 256 dimensional space choosing the bases vectors (=image components) such that most of the variability between images can be described by just a few components.

In the analysis, it will be apparent that the chips fall into three groups, where the Back Illuminated (BI) chips behave similarly, I0 and I2 do, and the remaining Front Illuminated (FI) chips form the last group. Most of the plots are thus done for just one or a few representative chips.

The gain also depends on the temperature of the Focal Plane, but that is beyond the scope of this study, all data here is from nominal cold observations with -120 $^\circ$C Focal Plane temperature.

All code for this analysis is available at \url{https://github.com/hamogu/dPHA/}.

\section{CALIBRATION DATA}
\begin{figure} [ht]
  \begin{center}
    \includegraphics[height=5cm]{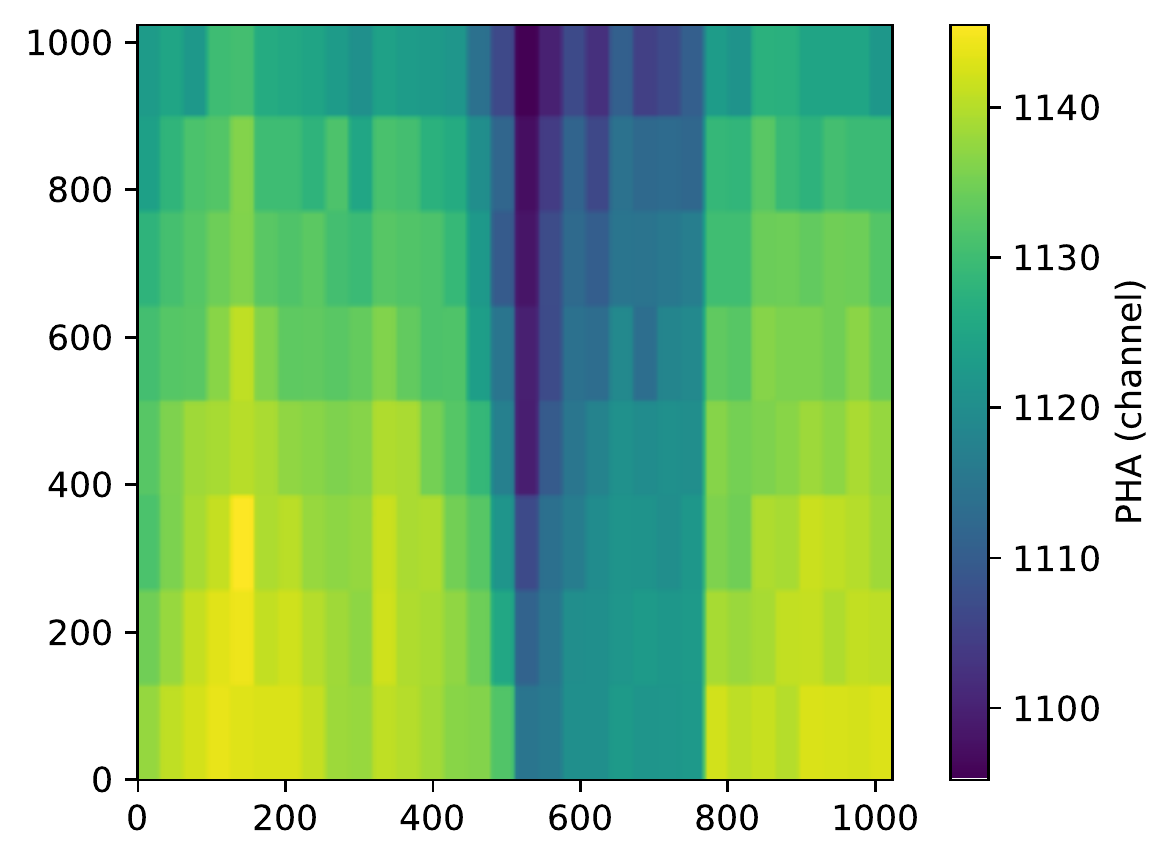}
  \end{center}
  \caption
      {Measured PHA values on the I3 chip in the Ti line in epoch 50. The signal is grouped into regions of $32 \times 128$ pixels each. The $x$ and $y$ axis are given in pixels.\label{fig:example}
}
\end{figure}

Data from the ECS has been used since Chandra's launch to calibrate the relation between PHA and photon energy. We reduce the data following the same established procedures. First, data is reprocessed with the standard \texttt{acis\_process\_events} procedure to correct for CTI, but not for the gain, since this is the quantity that we aim to establish. Events are split by chip and within each chip in regions of $32*128$~pixels, which gives 256 regions per chip. For each region, a spectrum is extracted and the position of the Al, Ti, and Mn K$\alpha$ lines are fit, see Figure~\ref{fig:example} for an example.
In this procedure the fits for each tile on the chip are independent. Yet, Figure~\ref{fig:example} shows that large scale structure exists. First, we can discern the node boundaries. In node 2 (nodes are counted starting at "0" -- $x$ pixel values 513-768) the PHA value of the Ti line peak is significantly lower than in the other three nodes. Furthermore, there is a clear trend with higher PHA values towards the bottom of the figure (low $y$ pixel values).

Smaller regions would allow a more fine-grained picture of the spatial dependence of the gain calibration. However, each region needs to a have a sufficient number of counts from the ECS to extract a spectrum and fit the centroid of the emission lines. If the number of counts is too low, then the fit might be unsuccessful or the statistical uncertainties of the fitted values might be too large. Even for the tiles chosen, this problem occurs since the count rate from the ECS declines as the $^{55}$Fe source decays over time. Until 2015, the gain calibration was performed every three months. Since 2016, data for six month is combined to ensure a sufficient count number to successfully fit the line position in the spectrum. Since 2020 region sizes were also increased from $32*32$ to the current $32*128$ pixels size to improve the fit statistics. Even with this adjustment, it happens that in some regions the number of counts is too low to fit the spectrum. In these cases, we fill in the missing values from the average of the neighboring tiles. No tiles ever needed filling on the S3 chip, while the outer S-array chips often require neighbor interpolation. A calibration observation program of increasing ECS exposure time has been implemented to provide a steady lower limit on the counts available for gain calibration.

\section{PRINCIPAL COMPONENT ANALYSIS}
Principal Component analysis (PCA) is a well known mathematical concept and a description can be found in standard text books. We are using the implementation from scikit-learn\cite{scikit-learn}. In short, we can look at the 256 regions in each image as a vector in 256-d space. PCA finds a new set of basis vectors and orders them such that the first vector in the list describes the direction with the largest variance between the input data points, the second vector the direction with the second most variance and so on. The new basis vectors always form an orthogonal basis (i.e.\ the PCA components are independent) and the process of calculating them is deterministic. (There are alternative implementations of PCA that make use of approximations to speed up the computation, but that is not a concern for the size of this dataset.) However, the new basis may not be complete. If only 215 datapoints are put into the PCA, the new basis will only have 215 vectors and thus not span the entire 256 dimensional space. PCA can be used as a tool for dimensionality and noise reduction. A simple example is a set of data points in 3D space that all fall onto a single plane with just some noise above and below the plane. In that case, the PCA will return two vectors that span up the plane and one that's perpendicular to it. The importance of this last component will be small. We can then decide to discard this component and only retain the remaining two, effectively forcing all points on the plane, which reduces the noise in one direction. At the end, those new noise-reduced points can be projected back into the original 3D space.

Similarly, we use PCA here to identify spatial components of the gain that vary together. For example, if the gain for the four nodes on the chip evolves with time, but in a different way, then we would get four components, one for each node. The images of those components would show the node in question and have a value of "0" for the others.

This description of a 256d vector discards the 2D structure of the image, no relation between individual components is assumed in the PCA. As we will see below, folding the 256-d vectors back into an $32 \times 8$ image will result in sensible 2D images. This is a useful sanity check.
For each chip, we run one PCA on data from all epochs and all three lines (Al, Ti, Mn) that we can measure from the ECS. The PCA returns components that describe most of the observed variance. For each point in time (epoch) and line (Al, Ti, Mn), we get a set of factors. Time evolution in the gain thus is described by the time series of these coefficients.
A technicality for the PCA is worth noting: The sign of the new bases vectors is arbitrary.

\begin{figure} [ht]
  \begin{center}
    \includegraphics[width=\textwidth]{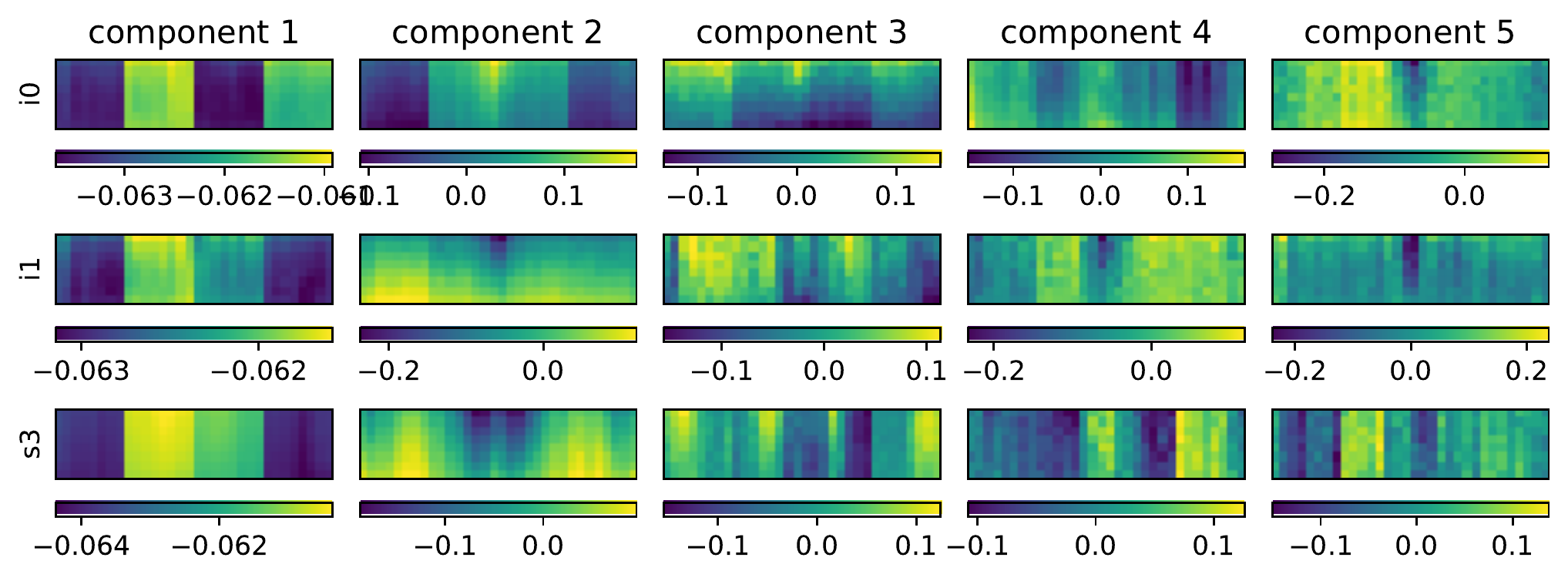}
  \end{center}
  \caption
      {Most important PCA components (left to right) for different chips (rows). Some similarities are seen between chips. \label{fig:components}}
\end{figure}

The most important PCA components are shown in Figure~\ref{fig:components} for different chips. We can notice similarities between the components on the chips. The most important component (left column) has a very small range of values, the difference between the minimum and maximum in the image of this component is only of order 0.02, while it is 0.2 to 0.5 for the other components. That means that all regions on the chip behave very similar when this component changes. We will see later that this component is essentially the linear scaling between photon energy and PHA. That proportionality differs slightly between nodes and ever more slightly within a node. The next components all have structure mostly along columns, with consistent changes from the top to the bottom of the chip
(recall that the sign is arbitrary here, since only the product of the scaling for this component $x_i$ and the 256-d vector of component values $C_i$ matter: $x_i \times C_i = -x_i \times -C_i$).

In most cases, spatial structure is clearly evident in these components, which is an indication that they have physical relevance. (Note that the PCA by itself does not know anything about the 2D structure. Each ``image'' is simply a 1D vector of 256 numbers.)

\begin{figure} [ht]
  \begin{center}
    \includegraphics[width=\textwidth]{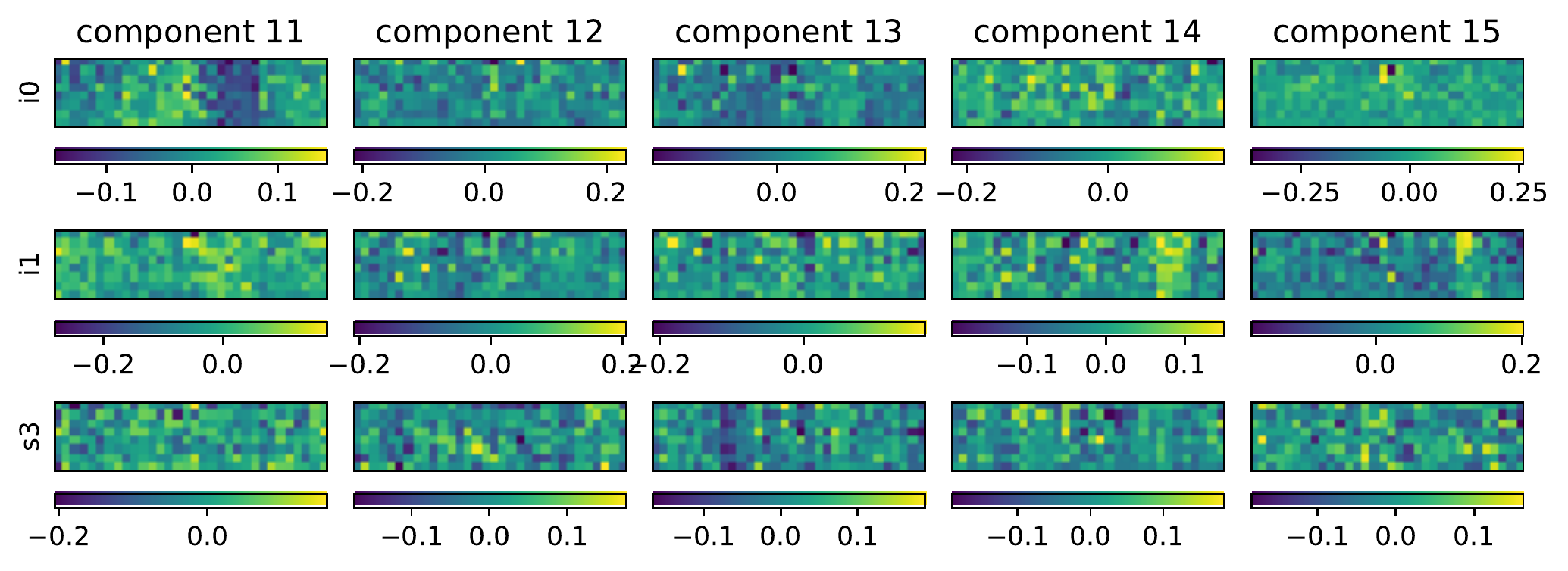}
  \end{center}
  \caption
      {The figure shows a few PCA components that explain less of the variance than those components shown in Figure~\ref{fig:components}. A few different components are shown left to right for different chips (rows). \label{fig:components_minor}}
\end{figure}

For comparison, Figure~\ref{fig:components_minor} shows some components of lesser importance. While these components are dominated by noise, some spatially consistent features are still present (e.g.\ the step between the third and forth node in the top left image).

\subsection{Explained variance and time evolution of scaling factors}

\begin{figure} [ht]
  \begin{center}
    \includegraphics[height=10cm]{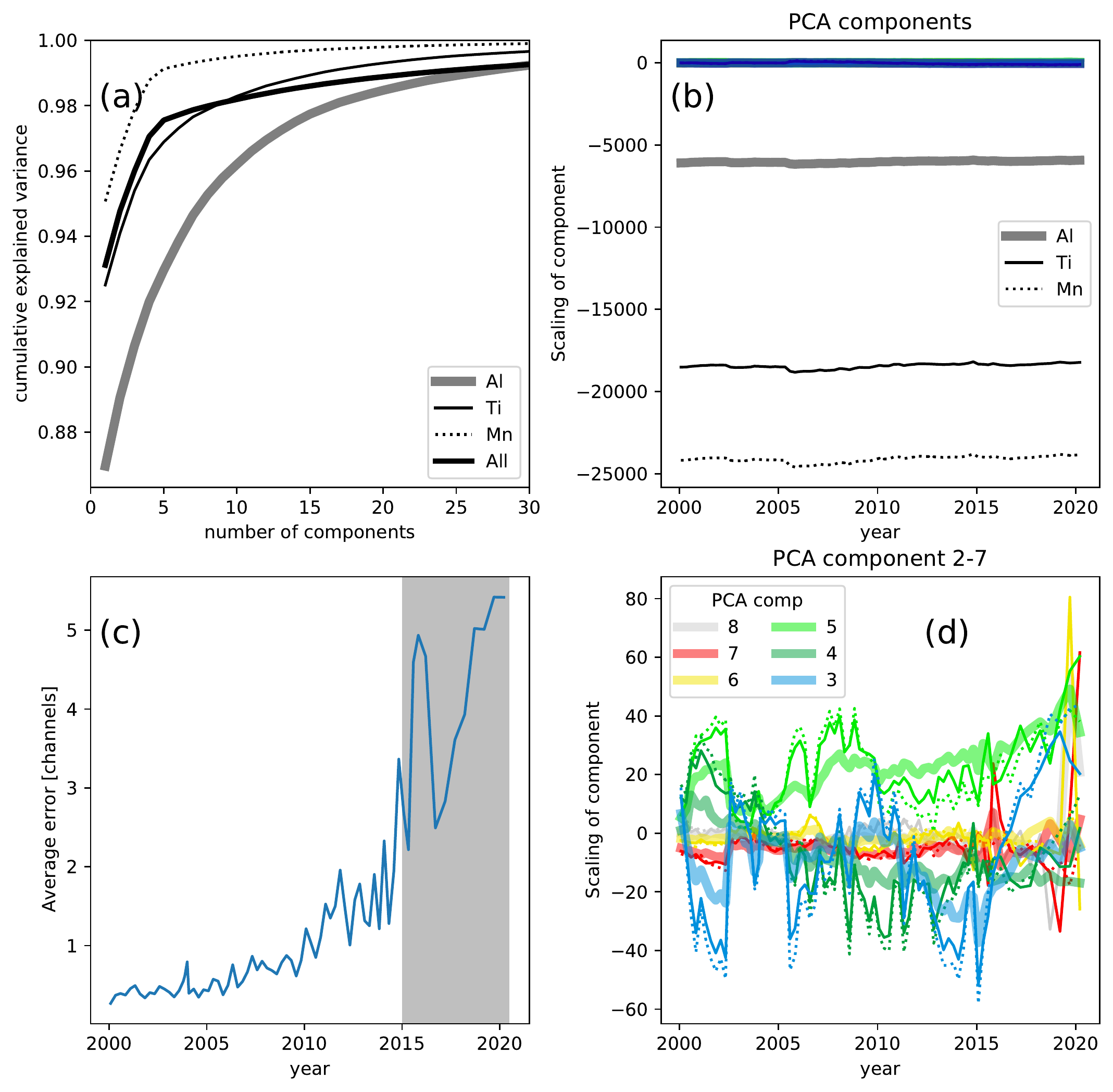}
  \end{center}
  \caption
      {PCA results for the I0 chip. \emph{(a):} Explained variance ratio (see text for details). Lines are shown for a PCA that is run for each spectral line independently or all three lines combined. In the combined case, just three PCA components are sufficient to explain 97\% of the observed variance. \emph{(b):} Scaling factor of PCA components. The first component is shown in black/grey with different line styles for different elements. Other component are shown in color, but coefficients are close to 0, so the cannot be seen individually in this plot. \emph{(c):} Average uncertainty in PHA channels of the fit of the line centroid. The radioactive decay of the $^{55}$Fe source leads to reduced signal over time and increasing uncertainties. Starting in 2015 (gray area), the calibration period has been increased to six month instead of three months. \emph{(d):} Same as (b), but zoomed in to see the less important PCA components.
      \label{fig:explvar}}
\end{figure}

The PCA orders components by importance, such that the first component explained the most of the data variance.  Figure~\ref{fig:explvar} (a) shows the explained variance for the I0 chip.
This is done four times: Once for Al, Ti, and Mn each (i.e.\ with 75 input epochs) and once using all the data for one chip at the same time ($3\times75=225$ input vectors). The more information is available, the better the PCA components will be. Since the gain maps for Al, Ti, and Mn are clearly similar, it makes sense to describe them with the same spatial components. As an example, for the I0 chip, 95\% of the variance in the Mn data can be explained with just one spatial component. The more components are used, the more of the variance will be explained. However, given that the input data contains noise, there is no need to explain everything, as this would result in fitting the noise. Instead, we are looking for the most important components only. In all chips, Al needs more components, but (not shown here directly) it is also the most noisy data. Note that for panel (a) in the plot only, we perform PCA not the raw PHA values, but subtract the epoch 0 fits. For the PCA of the full PHA values, the largest difference between the data of Al, Ti, and, Mn is that these lines are taken at different energies. A first component of the PCA takes care of this linear dependence and thus explains $>99.99\%$ of the variance. However, we need to decide how many component to look at beyond the simple proportionality with energy and thus it is more instructive to look at a PCA of the PHA values with linear energy dependence taken out.

The right column in Figure~\ref{fig:explvar} (panels b and d) shows the scaling factors for the first few PCA components over time. Different colors are for different components, different line styles for Al, Ti, and Mn. For example, the Ti image of a particular epoch can be written as scaling of component 1 multiplied by the spatial component 1 (Figure~\ref{fig:components}), plus the scaling of component 2 multiplied by component 2 and so on.

The first (the strongest) component is much more important than the next few, so this is the same plot without the first component. The scaling of the the PCA components for Al, Ti, and Mn behaves very similarly. This is explored in more detail below. These plots also give a first hint which components are physical and which start to describe the noise. Components that describe physical processes on the chip should have trends in time, because the values are correlated from one epoch to the next. On the other hand, components that just describe the noise pattern fluctuate around 0 and start to fluctuate more at later times when the noise in the data increases, e.g.\ components 6 or 7 for the I0 chip.

\subsection{Comparing chips}
\begin{figure} [ht]
  \begin{center}
    \includegraphics[width=\textwidth]{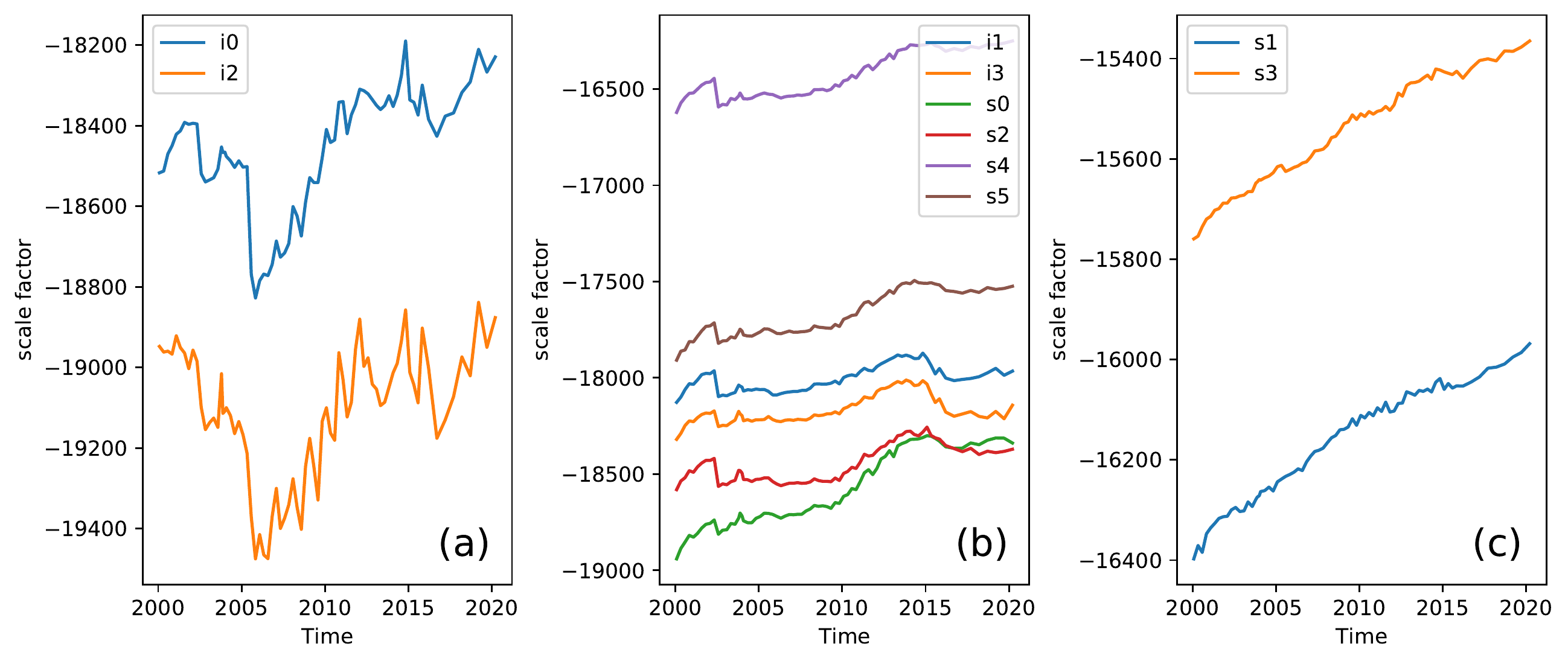}
  \end{center}
  \caption
      {Scale factor for the strongest component in the PCA for Ti. Chips are empirically sorted in groups with different temporal behaviour.
      \label{fig:comparechips}}
\end{figure}

Figure \ref{fig:comparechips} shows that chips fall into three groups with different behavior: For back-illuminated CCDs, the scale factor of the strongest component changes essentially linear with time (with a small offset in the first two years, see panel (c) of Figure \ref{fig:comparechips}). The front-illuminated chips show two different behaviors: While I2 developed an offset from I0 in the beginning, both chips have been developing in parallel since about 2003 (panel a). The remaining chips (I1, I3, S0, S2, S4, S5) form another group with very similar time evolution (panel b). In all cases, the scale factor moves up quickly between 2000 and 2003, when a sudden jump down occurs that bring some of the chips back down to 0. After that, this group develops largely in parallel, with the same offset. Only in recent years (after around 2015) do the scale factors diverge more.

\section{WHICH PCA COMPONENTS ARE SIGNIFICANT?}

Figure~\ref{fig:explvar} shows that the noise increases significantly over time (panel c). Panel d displays which PCA components contribute to the image over time. In all cases, the first few components are important, and they are important over the full time range, and there is significant coherent structure. In contrast, the higher components (e.g.\ 6, 7, 8 for I0) only turn up with a significant contribution in the last year or so and fluctuate widely. That indicates that these components are mostly fitting the noise. In the following we give three different lines of reasoning that only the first four or five PCA components matter.

\subsection{Time dependence of scaling factors}
\begin{figure} [ht]
  \begin{center}
    \includegraphics[height=5cm]{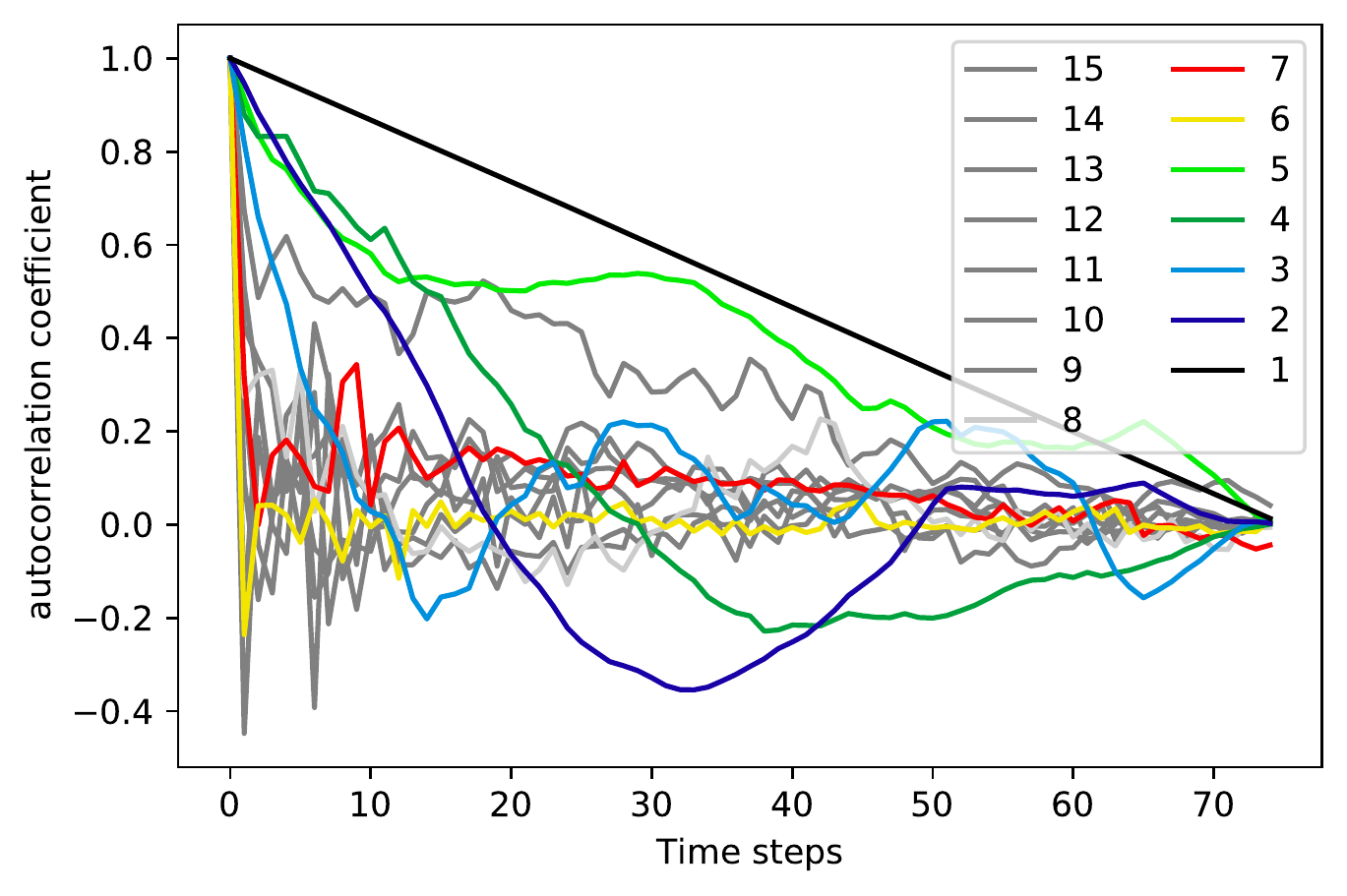}
  \end{center}
  \caption
      {Plot of the autocorrelation functions for the first 15 components in Ti for chip I0. Components 8-15 are all plotted in different shades of gray.
        \label{fig:time}}
\end{figure}

Figure~\ref{fig:time} plots the autocorrelation function for the time dependence for each component. Random noise is random in time and thus consecutive measurements are unrelated and the autocorrelation will drop close to 0 for time step 2. On the other hand, physical changes in the chip or the read-out do not appear one day and return to a previous state the next, so we expect that the scaling factors for PCA components that are physically relevant are correlated between different epochs.
In the figure, components 1-5 are different from random noise. Note that there are later components that have a long-time scale autocorrelation (e.g.\ one of the dark grey ones). These components likely also describe physical properties of the CCD, but they are less important than some of the noise-only components (components are sorted by the order of importance by the PCA) and, as we show in the next section are not consistent between different energies.

\subsection{Correlation between Al, Ti, and Mn lines}

\begin{figure} [ht]
  \begin{center}
    \includegraphics[height=5cm]{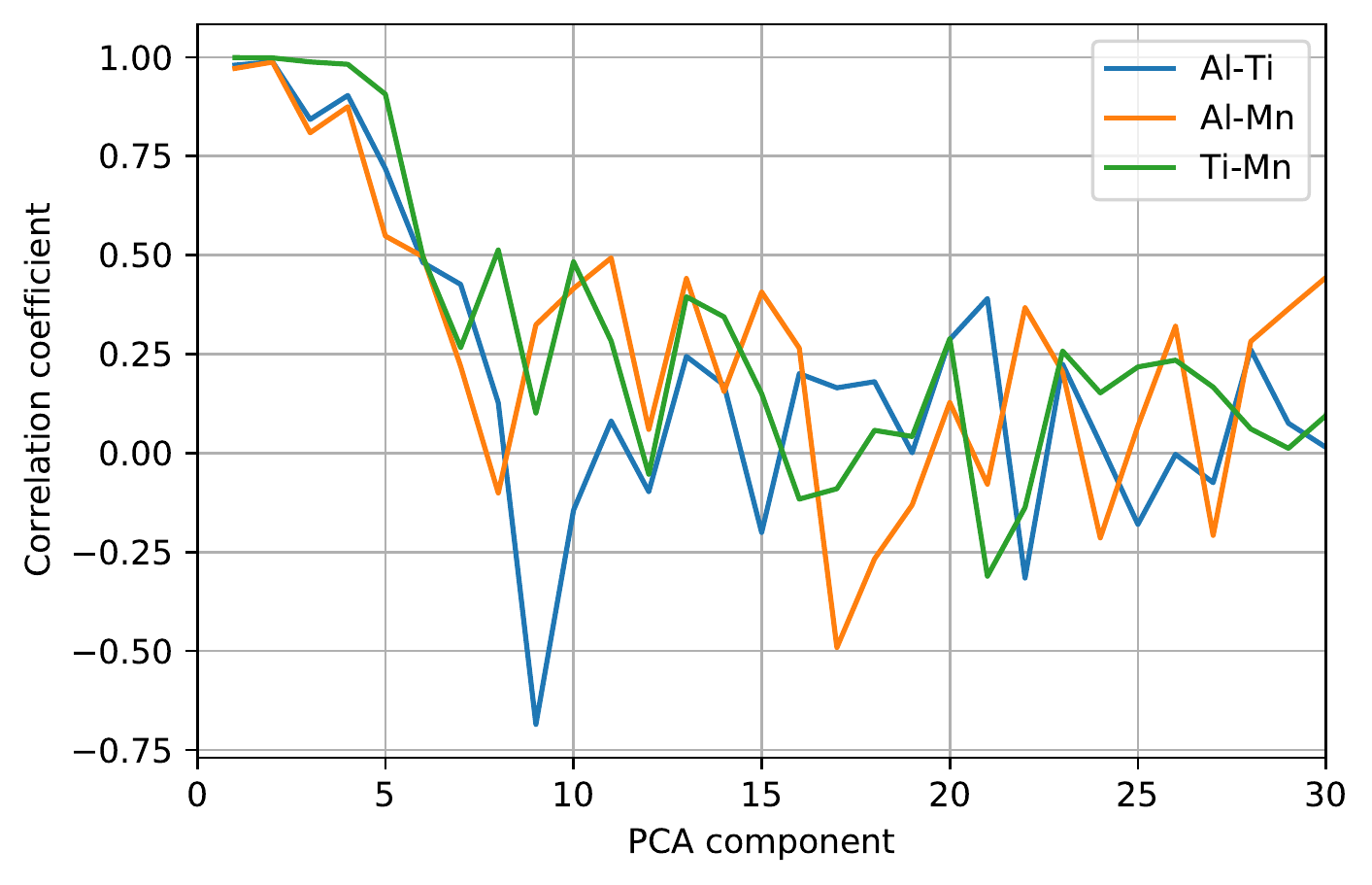}
  \end{center}
  \caption
      {Correlation between the Al, Ti, and Mn corefficients for chip I0.
        \label{fig:encorr}}
\end{figure}

Similarly, we can correlate the curves for Al, Ti, and Mn. Random noise is uncorrelated, while physical components should affect the gain in each line (Figure~\ref{fig:encorr}).

For each PCA component, we have three time series of scaling factors, one from the Al, Ti, and Mn lines each. While the gain values do have some energy dependence, physical changes in the chips should be seen in all three lines in some way and thus the three time series should be correlated. On the other hand, pure noise will be uncorrelated. The plots above show the correlation coefficients between the time series for Al, Ti, and Mn. The first three or four components show strong correlations and after that the coefficients are compatible with random noise again.

\subsection{$\chi^2$ test}
\begin{figure} [ht]
  \begin{center}
    \includegraphics[height=5cm]{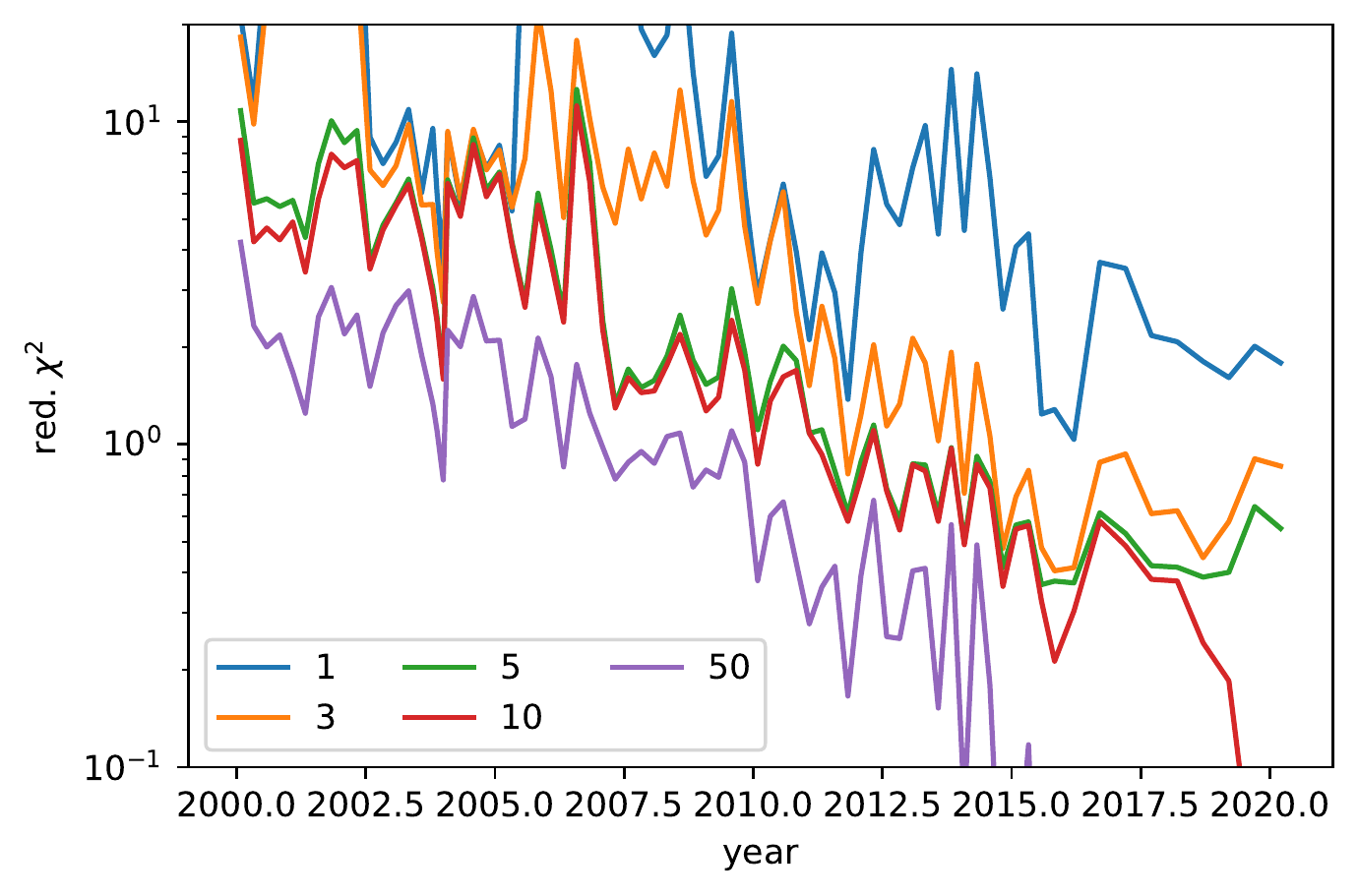}
  \end{center}
  \caption
      {Reduced $\chi^2$ when describing the Ti line on the I0 chip with a different number of PCA components.
        \label{fig:chi2}}
\end{figure}
For every measures line centroid in every tile, we know the uncertainty. Thus, we can describe the gain map with just the top PCA component and calculate the $\chi^2$-statistic. As we use more than one component, the description will be more accurate and thus the $\chi^2$ value will be lower; we can then see how many component are required until we obtain a satisfactory number for the reduced  $\chi^2$.
Figure~\ref{fig:chi2} shows the reduced $\chi^2$ for the I0 chip in each epoch, when using 1, 3, 5, 10, or 50 PCA components for each epoch. In general, the $\chi^2$ drops with time for any chosen $n$. We already showed above that the measurement uncertainty increases with time, as the intensity of the calibration source decreases. In itself, that is not surprising. While the shapes of the curves are similar, the scale differs. For most chips, several dozen components are necessary to reach  $\chi^2$ around one for early times, in other words, each region on the detector actually evolves different. However, with decreasing signal, less and less of that can the measured reliably, and at later times, after 2015 or so, good fits can often be obtained with just one or two spatial components. In several chips, there is a noticeable uptick in $\chi^2$ in 2019. This is possibly related to the count number becoming so low that a Gaussian error approximation is not appropriate any longer. Plots for Al and Mn are not shown here, but look similar.

This analysis indicates that from 2010 on all chips can be described by just four or five components, which is compatible with the analysis in the last two figures. However, at early times more components are needed to reach an acceptable $\chi^2$ value, which is somewhat surprising. If really more than five components are needed for the first 10 years (half the number of observations so far), one would expect that to show up in the autocorrelation of the correlation coefficients. Alternatively, the error which goes into the $\chi^2$ calculation could be underestimated.

\section{DISCUSSION OF PCA ANALYSIS}
While the gain maps in the beginning of the mission contain a lot of information about small scale changes, in the last few years, the noise has grown to the point where a satisfactory $\chi^2$ can be obtained by describing the  maps with just one (or a few) components. That means that the gain calibration products produced from fitting the observations in each region independently carry more noise than justified. Instead, it is preferable to fit the maps from about 2015 on as a linear combination of just one or a few components. That way, we have only 1 (or a few) free parameters, which can be determined much better than 256 independent regions.

For future calibrations, we can just fit one (or a few) parameters and construct the maps as a linear combination of the known most important (or the few most important) spatial components. Consequently, there is no need to illuminate the entire detector. Instead, we can place an astrophysical calibration source, such as Cas A, so that it illuminates only a few regions. If we use only one spatial component, any region on the detector will do. If we decided to use three or four spatial components, the source should be placed such that that fit as a good handle on the parameters, i.e.\ area where the spatial shape of the top few components is very similar should be avoided and area where the chosen components have large, and most importantly, different spatial gradients are ideal.

In particular the BI chips are well described with just one spatial component, which scales  almost linearly with time. The front-illuminated chips fall in two groups (I0/I2 are one group and the remaining chips the other) with similar temporal behaviour. Observations of one chip per group could be used to predict the behaviour of the others over short time periods. However, the correlation is not perfect and all chips need to be calibrated individually regularly.

\subsection{Does this method introduce new systematics?}
The PCA is fundamentally only a different way of looking at the same measurements. It still starts with fitting the peak position of the emission lines observed. Using the ECS as a data source, each of those fits still has the same systematics it had before. For example, if the position of the Ti line was unreliable because of unresolved line blends, and this introduced a systematic offset in the position of the Ti line peak, then the same systematic offset will also be part of the new basis vectors determined through PCA. So, the PCA does not remove those systematics, but there is also no reason to believe that it introduces new ones.

\subsection{How does solar activity impact the PCA?}
ACIS has operated through two full solar cycles already. As such, components in the gain maps that only appear during solar maximum are part of the input data to the PCA. If these components are important to explain the variance of all PHA maps taken, then they will be part of the top few PCA components with a coefficient that is large during a solar maximum and small during a solar minimum. Thus, when the next solar maximum comes up, the fits for these components will simply return a larger coefficient than they currently do.

\subsection{Testing if new PCA components are required}
ACIS experienced some sudden changes in the past and these are reflected by sudden jumps in the coefficients of each PCA component such as those seen in Figure~\ref{fig:explvar}, panel (d) around 2004 and 2006. The coefficients for component 3 and 5 suddenly drop close to 0 at this point, but then the same components rise in importance again in later years indicating that even in the event of sudden changes the same physical mechanisms are still responsible for gain changes. Of course, past performance over 20 years is no guarantee for future behavior, but at least a good indication for the typical development. We do not have to rely on this though: We can calculate the $\chi^2$ value for any new measurement of the gain and can use that to decide if the fit is satisfactory or if additional calibration observations that truly sample the entire chip with high S/N are required.

\section{CAS A AS CALIBRATION SOURCE}
While PCA can be used to de-noise the gain maps obtained from ECS measurements (because we need to determine only a few parameters instead of 256 independent measurements), the rapid decline of the ECS count rates as the $^{55}$Fe source decays makes is necessary to switch to using astrophysical calibration sources. Two problems with this approach are that astrophysical sources do not fill the entire field-of-view in Chandra with sufficient signal and that astrophysical sources do not provide constant emission lines at fixed energies. PCA can help with the first problem, because it allows us to reconstruct the gain map for the entire detector from measurements for only a few tiles. The second problem requires a careful choice of the calibration source and knowledge of its astrophysical properties. For example, for the bright supernova remnant Cas A, different emission regions display different ionization balances and kinematic motions, shifting the apparent line center\cite{2010ApJ...725.2038D,2012ApJ...746..130H}. Correction for this effect are outside of the scope of the current proceeding.

We now present an initial attempt (without corrections for astrophysical motion) of using Cas~A as a calibration source.
This is based on nine observations (ObsIDs 23257, 23258, 23259, 23260, 23261, 23262, 23263, 23264, 23265) of 2~ks exposure time each with Chandra. All nine observations were carried out on 2020-07-14 using the ACIS-I3 chip in standard FAINT mode and Cas~A was placed on different parts of the chip, since the source is too small to sample the entire field-of-view in a single observation. We combine data from all nine observations to obtain a ``baseline'' gain map. We then pick a single ObsID and treat this as our calibration observation. We fit data from just that one ObsID, fit a few PCA components to it, and evaluate the PCA components over the entire chip to build a predicted gain map. We compare that map to the ``baseline'' to assess the accuracy of this procedure.

\subsection{Measuring the channel where emission lines peak}
An X-ray spectrum of Cas A with the most important lines labeled can be found in e.g., \citenum{1994PASJ...46L.151H,2010ApJ...725.2038D,2012ApJ...746..130H}; the spectral plot is not reproduced here. We select a few bright lines that seem to be free from major contamination: Fe at 6.7~keT, Ca at 3.9~keV, Ar at 3.1~keV, S at 2.4~keV and Si at 1.9~keV. Except for Fe, the emission comes from He-like ions and is made up of the unresolved He-triplet emission. However, weaker, unresolved lines exist over the entire range of the spectrum. We fit the chosen lines with a very simple model, just a single Gaussian for the line, plus a pseudo-continuum powerlaw. First, we perform the fit on the full data integrated over the entire chip to select the position and width of the Gaussian as well has the slope of the power-law. Then, we fix the width of the Gaussian and use its position and the slope of the powerlaw as starting values for the fits that we do on individual tiles. Having few parameters and starting the fit very close to the best-fit position speeds up the fitting process and reduces the risk that the fit runs into local minima.

\subsection{PCA for a Cas A observation}
\begin{figure} [ht]
  \begin{center}
    \includegraphics[width=\textwidth]{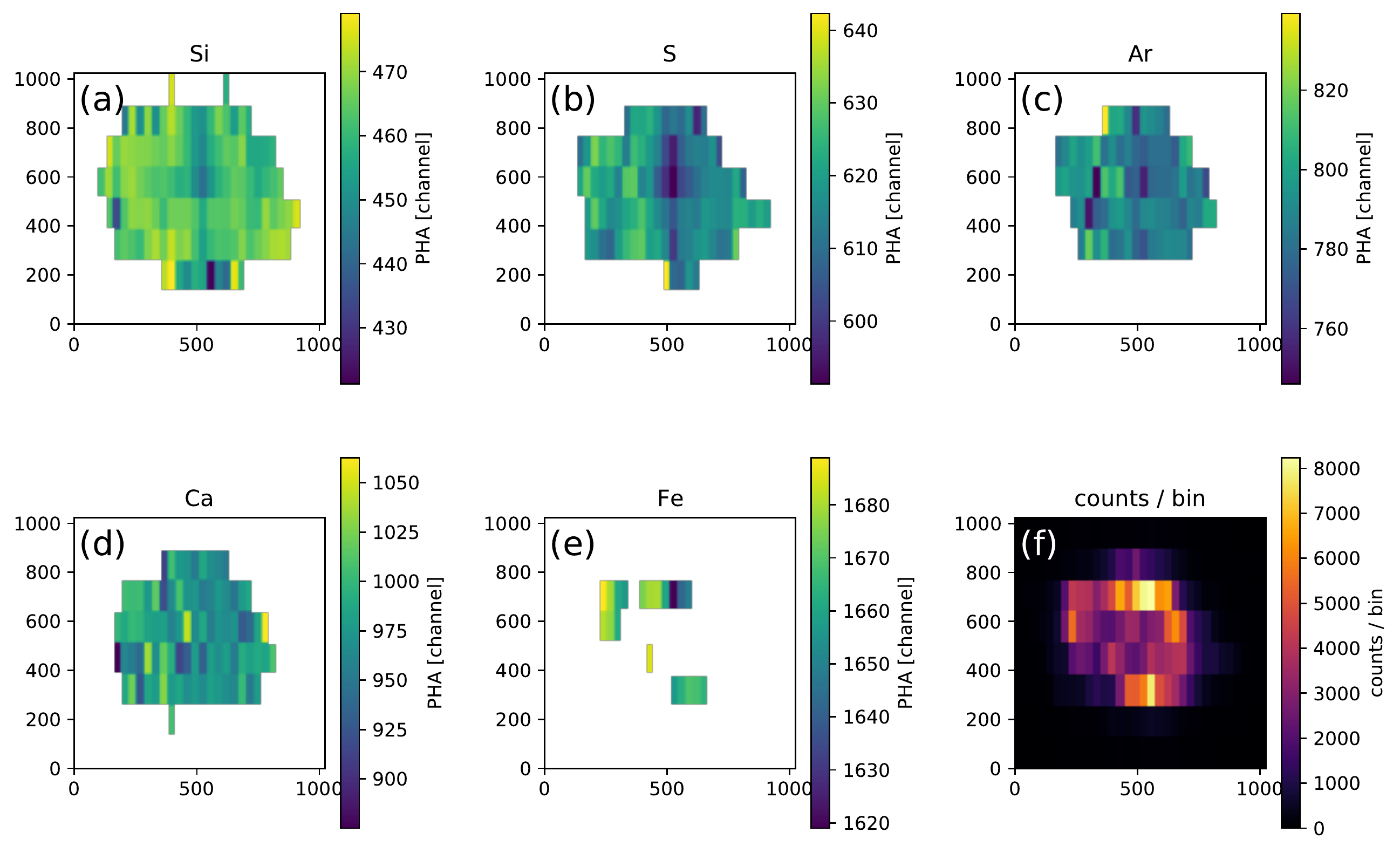}
  \end{center}
  \caption
      {\emph{(a-e):} Peak PHA channel for emission lines form different elements in Chandra observations of Cas A (ObsID 23261) \emph{(f):} Total number of countsdetected photons in each spatial bin (tile) on the detector. Fits for each lines are done using only photons with an energy close to the expected line center, but this panel gives a good impression which regions of the detector are exposed well.
        \label{fig:CasAlines}}
\end{figure}

Figure~\ref{fig:CasAlines} shows the fitted peak position of selected lines for ObsID 23261. Regions with insufficient counts for a fit are left white in the image. The signal in Si, which is the strongest emission line, is best. Regions with $<50$~counts are too noisy to be useful, so those are skipped.
In general, we can see the same spatial pattern that emerges from the ECS data, where some regions are clearly more noisy than others simply because there are fewer counts.

\begin{figure} [ht]
  \begin{center}
    \includegraphics[width=\textwidth]{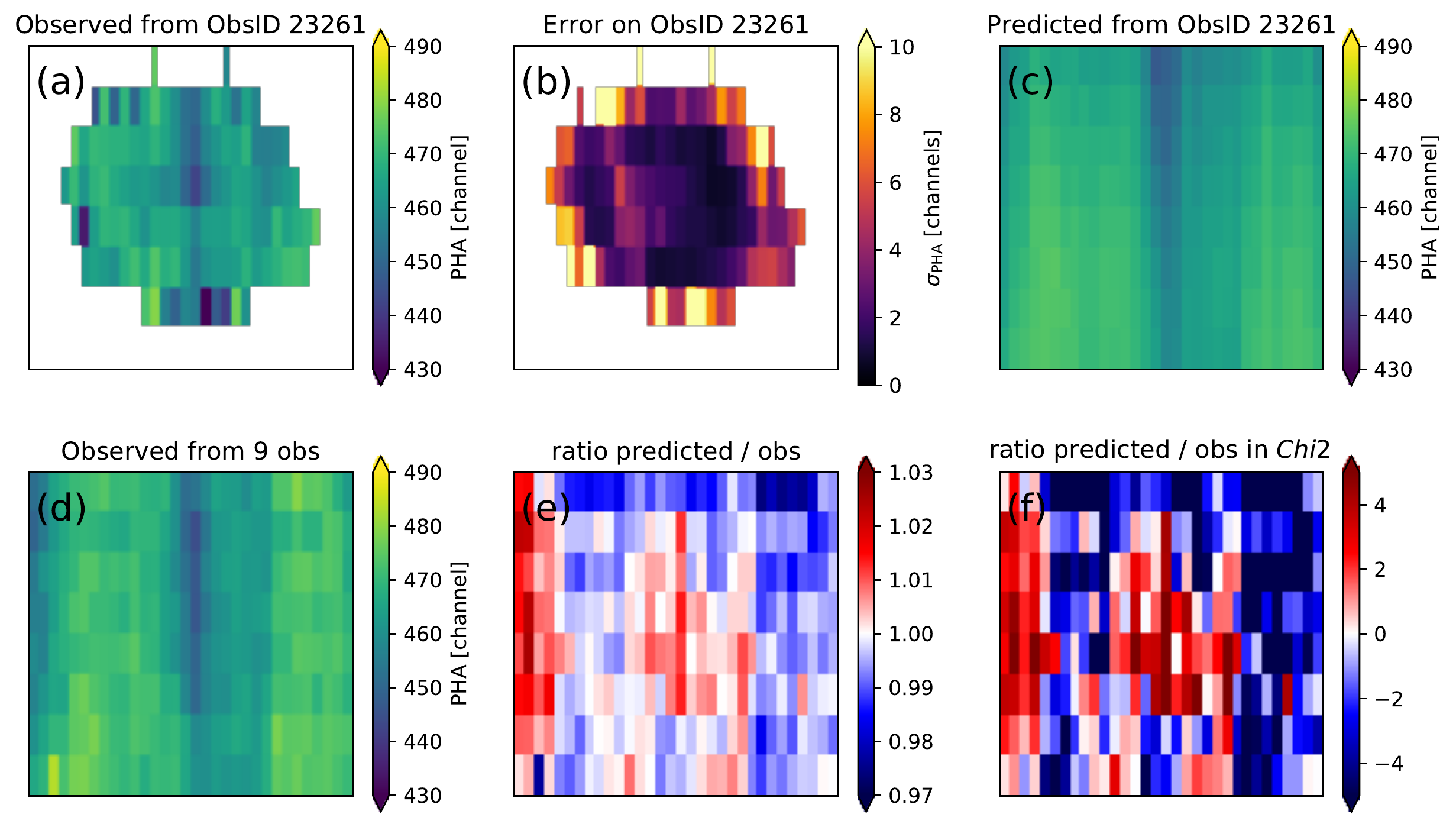}
  \end{center}
  \caption
      {\emph{(a):} Peak position of Si line for ObsID 23261 \emph{(b):} Uncertainty on peak position for ObsID 23261 \emph{(c):} predicted peak position from a fit of the top four PCA components \emph{(d):} Peak position of Si line for all nine ObsIDS combined. \emph{(e):} Ratio of predicted (from the PCA fitted to just one ObsID) and the best-known peak position of the Si line \emph{(f):} Like (e), but now divided by the statistical error on the best-known position in each tile.        \label{fig:Si23261}}
\end{figure}

Figure~\ref{fig:Si23261} shows in some detail how we can fit the Cas A data with the top few PCA components. To evaluate the validity of the method, we need to compare the predicted gain in each tile with the true gain for the Si line energy. Unfortunately, we do not have a perfect measure of the latter quantity. The ECS data is noisy and also does not cover the exact energy of the Si line. Thus, we merge the event lists from all nine Cas A observations and treat this set as the baseline. Some caution is warranted: In Cas A kinematic motions are large enough to disturb the measurement (plasma moves up to 1\% of the speed of light, shifting the peak of the line position by up to 1\%) and even in the co-added data, the S/N is not ideal close to the chip edges, simply because those edges are not illuminated very well given the coordinates chosen for the observations.

However, in general, we can see broad agreement between the maps of the peak positions predicted from just a single ObsID and the map measured from a combination of all nine ObsIDs. Systematic differences are seen on the left edge of the chip (red columns in panel e) and in the top row (blue in panel e). Both are regions that are not covered in ObsID 23261. We know from the PCA components however, that some components parameterize a dependence of the gain from the top to the bottom, presumably caused by CTI. Since ObsID 23261 covers mostly the middle of the chip, those components cannot be well constrained in the fit. Figure~\ref{fig:Si23261} displays fits for the Si line.

The signal from Cas A is weaker at higher energies and consequently the noise is much larger for Ar or Ca than for Si. There is still enough data for fit four PCA components to these lines, but the comparison to the increasingly noisy data from the nine combined observations becomes less meaningful. In this case, one can see how the use of PCA as a noise reduction tool is useful.

\subsection{How should calibration observations be set up?}
Going forward, we can be more deliberate in where exactly we place Cas A on the detector. As we saw above, this matters as regions not sampled may not be fit well. For example, the node-to-node differences cannot be accurately predicted unless there is at least some data on each node.

\begin{figure} [ht]
  \begin{center}
    \includegraphics[height=5cm]{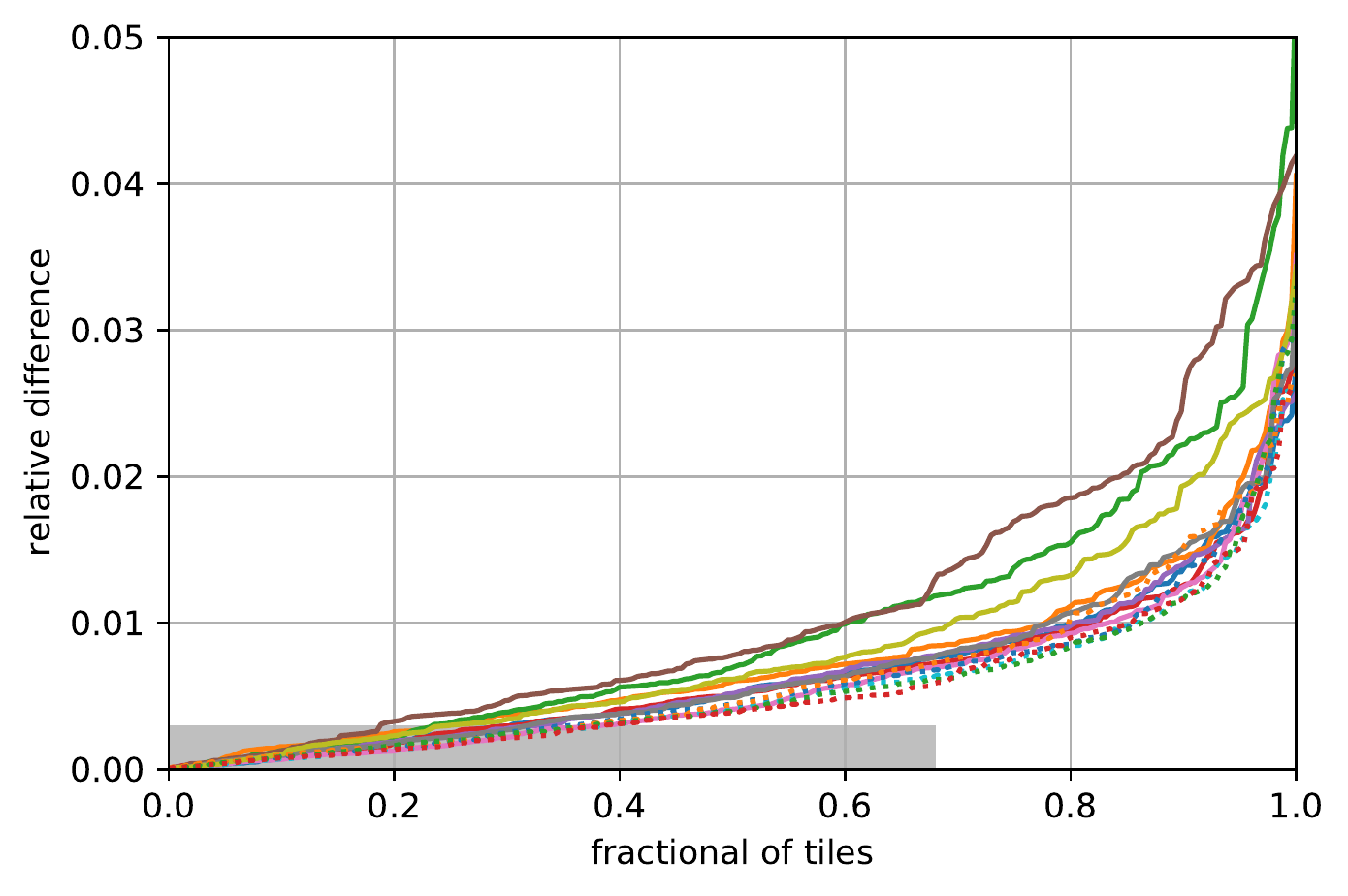}
  \end{center}
  \caption
      {Cumulative histogram of the ratio between the gain map, predicted using just one or two ObsIDs and PCA and the full gain map from all nine observations. Solid lines are from a single ObsID, dotted lines from combining two ObsIDs.
        \label{fig:final}}
\end{figure}

To further quantify this, we repeat the analysis in Figure~\ref{fig:Si23261} for other ObsIDs, where Cas A was located at a different position on the detector or for combinations of ObsIDs. We do not reproduce Figure~\ref{fig:Si23261} for all other ObsIDs in this proceeding, instead, we just show a histogram of the absolute values from panel (e). This captures the important information on how well the gain map constructed from just one ObsID compares to the map that we adopted a baseline for this analysis (a purely observed map, with no application of PCA, using all nine ObsIDs). In Figure~\ref{fig:final} we show these distributions for all nine ObsIDs (solid lines) and for some combination of two ObsIDs (dotted lines). In all cases, some tiles match better than others, but a large fraction of tiles (between 60\% and 80\%) matches within 1\%. The gray box indicates the way in which Chandra calibration has measured performance of the algorithm in the past: At least 68\% of the tiles should have the gain matched within 0.3\%.

The most important result from this exercise is that we can reach an accuracy of about 0.6\% for 68\% of the tiles. To do so, we need to select a placement of Cas A roughly in the middle of the chip, such that all nodes are covered, or combine several observations, such that the combined observation covers all nodes.

There are a few consistent patterns. In particular, all models overpredict the left edge of the chip. This is true even when the prediction is made from the combined data of all observations. We repeated these fits using a much larger number of PCA components and still observed this effect. A possible explanation is that the chip has evolved recently in a way not seen in the ECS data and that this spatial component is therefore not represented in the PCA components. However, ECS data taken both before and after do not show the deep dip on the left side of the chip in the Al line, which is close in energy to the Si line shown here. In the ECS data, the gains drops off on the left chip edge, but consistent with the slight drop seen in the predictions (from about 470 in the middle of the first node to 460 on the edge of the chip), while the data shows a drop from 470 to 450. This feature drives the high end of the cumulative differences.

We suspect that this is an astrophysical effect, i.e.\ that the Si line in this region is systematically shifted due to kinematic motion of the gas. Further work is required to quantify and remove this effect. In the middle of the chip, all nine observations contribute counts which come from different regions of Cas A. Some of that will average out the most extreme velocities. However, on the chip edges only one or two observations contribute, so astrophysical motion does not average out over many observations.

\subsection{How good is the calibration?}
None of the curves in Figure~\ref{fig:final} appear to match the Chandra requirement (gray box). However, several caveats have to be kept in mind: The input data is not yet corrected for astrophysical kinematics (that is ongoing work), which may remove some systematics. Also, we use fits from all nine observations as the ``baseline''. In some parts of the chip, even the combined data is noisy. Thus, even if the PCA would predict the gain map perfectly, we would still see deviations, simply because of the noise in the baseline. We also note, that the Chandra goal of ``68\% of the tiles match better than 0.3\%'' technically allows for arbitrarily large deviations on 32\% of the tiles. That could be catastrophic for a science user looking at an object located on one of those tiles! In practice, the calibration team will have this in mind when selecting the calibration strategy.

\section{SUMMARY AND CONCLUSION}
We present a PCA analysis of archival Chandra calibration gain maps. We find that, especially in recent years with the calibration source decaying, the gain maps for all chips can be described by just a few spatial components. This allows us to predict the gain maps for the full chip, even if only some tiles on the chip have sufficient calibration data. This procedure allows us to extend the useful life of the ECS, because lower count rates can still be used for calibration, and it enables us to make use of astrophysical calibration sources that do not fill the field-of-view. We present observations of Cas~A, a likely future calibration source. We investigate its properties and find that there are likely astrophysical effects that need to be corrected to improve the accuracy of the calibration (work in progress). With just the procedures presented here, we can perform a calibration that brings 68\% of the tiles within 0.6\% of our adopted baseline gain value (see discussion above on the assumption going into that baseline). While this is below the pre-launch accuracy requirements, it is much better than not having a gain calibration at all at the end of the useful live of the ECS and requires only a modest (2~ks per chip) amount of observing time.

\acknowledgments
Support
for this work was provided in part through NASA grant NNX17AG43G and
Smithsonian Astrophysical Observatory (SAO) contract SV3-73016 to MIT
for support of the {\em Chandra} X-Ray Center (CXC), which is operated
by SAO for and on behalf of NASA under contract NAS8-03060.
The
analysis uses Astropy, a community-developed core Python
package for Astronomy\cite{astropy1,astropy2}, numpy\cite{numpy}, scikit-learn\cite{scikit-learn} and
IPython\cite{IPython}. Displays are done with
matplotlib\cite{matplotlib} and bokeh\cite{bokeh}.
\bibliography{report} 
\bibliographystyle{spiebib}

\end{document}